\documentclass[aps,pra,superscriptaddress,amsmath,amssymb,preprintnumbers,floatfix,showpacs,11pt]{revtex4}

\usepackage{amssymb}
\usepackage{epsfig}

\begin{document}
\title{ Squeezing evolution with non-dissipative  $SU(1,1)$ systems }
\author{ Faisal A. A. El-Orany}
\email{el_orany@hotmail.com; faisal.orany@mimos.my }
 \affiliation{Department of Mathematics  and computer Science,
Faculty of Science, Suez Canal University 41522,
 Ismailia, Egypt;
 }

 \author{ S. S. Hassan}
\email{Shoukryhassan@hotmail.com}
 \affiliation{ University of Bahrain, College of Science, Mathematics Department,
 P.O. Box 32038, Bahrain}

\author{ and M. Sebawe Abdalla}

 \affiliation{Mathematics Department, College of Science, King Saud
University, P.O. Box 2455, Riyadh 11451, Saudi Arabia}

\date{\today}

\begin{abstract}
We investigate the squeezed regions in  the phase plane for
non-dissipative dynamical systems controlled by $SU(1,1)$ Lie
algebra.
   We analyze such study
for the two $SU(1,1)$ generalized coherent states, namely, the
Perelomov coherent state (PCS)
 and the Barut-Girardello Coherent state (BGCS).

\end{abstract}

  \maketitle

%
\section{Introduction}
The $SU(1,1)$ algebra (together with the $SU(2)$ algebra) has been much
utilised in the theoretical study of the non-classical aspects of light in various
quantum optical systems. For example: (i) The single-mode and the
two-mode squeezed vacuum states of light, each with different bosonic algebraic
realisation, are Perelomov-type $SU(1,1)$  coherent states \cite{per1}.
(ii) The even and odd coherent states of light \cite{(2),{(3)}} are one-mode
bosonic realisation of the  $SU(1,1)$ Lie algebra. (iii) The generalized
 Barut-Girardello coherent state is two-mode bosonic algebraic
 realisation of the  $SU(1,1)$ Lie algebra \cite{baru}: such states
were used in the theoretical description of coherent pion production in high
energy collision processes (see \cite{(5)} and references therein)
and within the framework of field theory in connection with "charged coherent
state" representation for an Abelian charged massless boson field in the
one-dimensional \cite{(6)} and infinitely many degrees of freedom \cite{(7)}
cases. Within the context of the field of quantum optics, Agarwal \cite{(8)}
introduced the name "pair coherent states" to such generalised coherent states.
(iv) The $SU(1,1)$ Lie algebra appears as the dynamical group of the quantum harmonic
oscillator (, e.g. \cite{(9)}). On the experimental side, optical interferometers
like four-wave mixers \cite{(10),{(11)}} are characterised by
$SU(1,1)$ Lie algebra (for different schemes of generating the $SU(1,1)$
states see \cite{(12),{(13)}} and references therein). Study of squeezing properties
associated with the $SU(1,1)$ (as well $SU(2)$) states aims mainly at the
reduction of the quantum noise in the act of measurements in fields like
spectroscopy \cite{(14)} and interferometry \cite{(15)}.

In the present
paper we study the time evolution of the squeezed regions associated with a
non-dissipative  model Hamiltonian which is presented as the linear
combination of  the generators
of the $SU(1,1)$ Lie algebra. For initial conditions, we consider the two
$SU(1,1)$ generalised coherent states, namely,  the Perelomov coherent state
(PCS)  and the Barut-Girardello Coherent state (BGCS). The present study
complements  similar investigation for the $SU(2)$ Lie algebraic representation
state \cite{hassan1}.
Nevertheless, it is worth referring to \cite{(20)} in which $SU(1,1)$
squeezing has been analysed for the $SU(1,1)$ generalised coherent
states when they are evolving in the
nonabsorbing nonlinear media modelled as anharmonic oscillator.
The dynamics related to such system is simpler than that in the present
paper, as we shall see in the following section.
Further in \cite{(20)} the author shows for the initial $SU(1,1)$ squeezed
states (, e.g. the PCS) that the system can destroy the initial squeezing
as well as can generate squeezing in the unsqueezed quadrature much
greater than that in the initial squeezed conjugate quadrature.
Also for the initial $SU(1,1)$ unsqueezed states (, e.g. the BGCS)
maximum squeezing can be
observed {\it only}
when the initial intensity of the field is very large.

The paper is organized in the following sequences:
In section $2$ we give the model Hamiltonian, the  solution of the
associated Heisenberg equations of motion and the basic relations and
equations which will be used in the paper.
In sections $3$ and $4$ we discuss the evolution of the squeezed regions
for  the PCS and BGCS state, respectively. The results are summarized
in section $5$.

\section{Model Hamiltonian and operator solutions}
In this section  we give the basic relations and equations, which will be mainly used
throughout the paper. The relations include the model Hamiltonian and its  associated
operators solutions,
the definitions of both the Perelomov $SU(1,1)$ coherent state (PCS) \cite{per1},
 the Barut-Girardello coherent state (BGCS) \cite{baru} and
 $SU(1,1)$ squeezing.

Now we consider a non-dissipative  Hamiltonian model of the form
(in units of $\hbar=1$),
\begin{equation}
\hat{H}=2\omega \hat{k}_{z}+\lambda (\hat{k}_{+}+\hat{k}_{-}),\label{f1}
\end{equation}
where $\hat{k}_{z}$ is a hermitian operator,
$\hat{k}_{+}=(\hat{k}_{-})^{\dagger}$ are raising (lowering) operators,
 $\lambda$ is a suitable coupling parameter and $\omega$ is a frequency
specifying the model.
For the $SU(1,1)$ Lie algebra the operators
$\hat{k}_{\pm,z}$
satisfy the following
commutators:
\begin{equation}
[\hat{k}_{+},\hat{k}_{-}]=-2\hat{k}_{z},\quad
[\hat{k}_{z},\hat{k}_{\pm}]=\pm \hat{k}_{\pm}.\label{i1}
\end{equation}
Note the Hamiltonian (\ref{f1}) represents either a degenerate or non-degenerate
parametric amplifier (, e.g. \cite{(17)}) according to the single mode boson
representation
$\hat{k}_{z}=\frac{1}{2}(\hat{a}^{\dagger}\hat{a}+\frac{1}{2}),
\hat{k}_{+}=\frac{1}{2}\hat{a}^{\dagger 2}$ or the two-mode boson representation
$\hat{k}_{z}=\frac{1}{2}(\hat{a}^{\dagger}\hat{a}+\hat{b}^{\dagger}\hat{b}+1),
\hat{k}_{+}=\hat{a}^{\dagger}
\hat{b}^{\dagger}$, respectively. In both cases the operators
$\hat{k}_{z}, \hat{k}_{\pm}$ obey the $SU(1,1)$ Lie algebra in (\ref{i1}).
For convenience  we define the following operators:
\begin{equation}
\hat{k}_{x}=\frac{1}{2}(\hat{k}_{+}+\hat{k}_{-}), \qquad
\hat{k}_{y}=\frac{1}{2i}(\hat{k}_{+}-\hat{k}_{-}), \label{f2}
\end{equation}
where the set
$\{\hat{k}_{x},\hat{k}_{y},\hat{k}_{z}\}$
 satisfies the following commutation rules
\begin{equation}
[\hat{k}_{x},\hat{k}_{y}]=-i\hat{k}_{z},\quad
[\hat{k}_{y},\hat{k}_{z}]=i \hat{k}_{x}, \quad
[\hat{k}_{z},\hat{k}_{x}]=i \hat{k}_{y}
.\label{i2}
\end{equation}
The associated Heisenberg uncertainty relation
  regarding  the first
commutator in (\ref{i2}) takes the form

\begin{equation}
\langle (\Delta \hat{k}_{x} )^{2} \rangle\langle ( \Delta \hat{k}_{y})^{2}\rangle
 \geq \frac{1}{4} | \langle
\hat{k}_{z}\rangle |^{2},\label{i3}
\end{equation}
where $\langle (\Delta \hat{k}_{j} )^{2} \rangle=
\langle \hat{k}_{j}^{2} \rangle-\langle \hat{k}_{j} \rangle^{2}$.
 To measure  squeezing, we define the functions
\begin{equation}
F_{j}=\frac{\langle (\Delta \hat{k}_{j} )^{2}\rangle- \frac{1}{2} | \langle \hat{k}_{z} \rangle |
}{
\frac{1}{2} | \langle \hat{k}_{z}\rangle |}, \qquad \qquad j=x,y.\label{i4}
\end{equation}
Squeezing (reduction) in the fluctuation of the $\hat{k}_{x}$- or
$\hat{k}_{y}$-components occurs if $F_{x}<1$ or $F_{y}<1$, respectively,
and
maximum  squeezing is reached  when
$F_{x}=-1$ or $F_{y}=-1$.

Now, the evolution of the operators
$\{\hat{k}_{x},\hat{k}_{y},\hat{k}_{z}\}$ according to the Hamiltonian
(\ref{f1}) and the relations (\ref{f2}), (\ref{i2}) are obtained by solving
the corresponding Heisenberg equations of motion, which read
\cite{hassan1,{sebaw1}}

\begin{eqnarray}
\begin{array}{lr}
\hat{k}_{x}(t)=R_{1}(t) \hat{k}_{x}(0)-J(t)\hat{k}_{y}(0)-S(t)\hat{k}_{z}(0)
\\\\
\hat{k}_{y}(t)=J(t)  \hat{k}_{x}(0)+R_{2}(t)\hat{k}_{y}(0)+V(t)\hat{k}_{z}(0)
\\\\
\hat{k}_{z}(t)=S(t)  \hat{k}_{x}(0)+V(t)\hat{k}_{y}(0)+R_{3}(t)\hat{k}_{z}(0),
\label{i5}
\end{array}
\end{eqnarray}
where  the $c$-number time-dependent coefficients have the forms
\begin{eqnarray}
\begin{array}{lr}
 R_{1}(t)=\cos(2gt)-\frac{2\lambda^{2}}{g^{2}}\sin^{2}(gt), \quad
R_{2}(t)=\cos(2gt),
\\\\
R_{3}(t)=\cos(2gt)+\frac{2\omega^{2}}{g^{2}}\sin^{2}(gt),
\quad
J(t)=
\frac{\omega}{g}\sin(2gt),
\\\\
S(t)=
\frac{2\omega\lambda}{g^{2}}\sin^{2}(gt),\quad
V(t)=
\frac{\lambda}{g}\sin(2gt),
\label{i6}
\end{array}
\end{eqnarray}
and $g=\sqrt{\omega^{2}-\lambda^{2}}$. It is clear that when $\omega
>\lambda$ the coefficients in (\ref{i5}) are periodic in the scaled time
$(gt)$ with period $2\pi$.

For convenience we give some remarks on
the Hamiltonian of the nonabsorbing nonlinear media, which can be
modelled as anharmonic oscillator. This
Hamiltonian--in the framework of $SU(1,1)$ Lie algebra generators--takes the form
\cite{(20)}:
\begin{equation}
\hat{H}=\omega \hat{k}_{z}+\lambda \hat{k}_{+}\hat{k}_{-}, \label{ff1}
\end{equation}
where  the notations have the same meaning as given above.
In the bosonic language  (\ref{ff1})  represents
 the Kerr media Hamiltonian.
For (\ref{ff1}) one can easily show  that  $\hat{k}_{z}$ and
$\hat{k}_{+}\hat{k}_{-}$ are constants of motion, and the solutions of
Heisenberg equations  are
\begin{equation}
 \hat{k}_{z}(t)= \hat{k}_{z}(0), \quad \hat{k}_{+}(t)
= \hat{k}_{+}(0)\exp[it(\omega +2\lambda \hat{k}_{z})]. \label{ff2}
\end{equation}
It is obvious  that  the
interaction Hamiltonian (\ref{ff1}) provides an overall phase factor in
the evolution operators $\hat{k}_{\pm}(t)$, i.e. it preserves the photon statistics of the
system. Keeping this in mind and  comparing
 (\ref{i5}) and (\ref{ff2}) one can conclude
that the dynamics associated with (\ref{ff1})  is so simple compared to that
of (\ref{f1}).

We proceed by considering  two types of  $SU(1,1)$ states, namely,
the Perelomov $SU(1,1)$ coherent state (PCS) \cite{per1}
and the Barut-Girardello coherent state (BGCS) \cite{baru}.

The PCS is defined as
\begin{equation}
|\xi ;k\rangle =(1-|\xi |^{2})^{k}\sum_{m=0}^{\infty
}\sqrt{\frac{\Gamma (m+2k)}{m!\Gamma (2k)}}\xi ^{m}|m;k\rangle, \label{i7}
\end{equation}
 where $\xi =-\tanh (\frac{r}{2})\exp (-i\Phi)$, with $
|\xi|\in (0,1),\quad r\in (-\infty,\infty),\quad \Phi\in (0,2\pi)$,
$\Gamma $ stands for Gamma function and $k$ is called Bargmann index
($k(k-1)$ is the eigenvalue of the Casimir operator
$\hat{C}=\hat{k}_{z}-\frac{1}{2}(
                   \hat{k}_{+}\hat{k}_{-}+\hat{k}_{-} \hat{k}_{+})$).
For $k=1/4(3/4)$ the PCS is  the even (odd) parity
coherent  state \cite{(2),{(3)}}.

The BGCS is defined as

\begin{equation}
 |Z;k\rangle =\sqrt{\frac{|Z|^{2k-1}}
{I_{2k-1}(2|Z|)}}\sum_{m=0}^{\infty}
\frac{Z^{m}}{\sqrt{m!\Gamma (m+2k)}}|m;k\rangle ,\label{i8}
\end{equation}
where $I_{k}(..)$ is the modified Bessel function of order
$k$, $Z=|Z|\exp(i\Phi),\quad \Phi\in (0,2\pi)$ and $k$ is the Bargmann index. It is worth
mentioning that these states are the eigenstate of $\hat{k}_{-}$, i.e.
$\hat{k}_{-}|Z;k\rangle =Z|Z;k\rangle $.

The explicit expressions for the expectation value of the arbitrary moments
$\langle \hat{k}^{l}_{-}(0)\hat{k}^{m}_{z}(0)\hat{k}^{n}_{+}(0)\rangle$
in both the PCS and the BGCS are given in
 \cite{(19)}.

For $t>0$ and as
 the interaction is turned on the energy exchange
between the
interacting subsystems starts to play a role and thus the shape of the squeezed
regions (in the complex plane)  depends on the ratio $(\lambda/\omega)$.
For this reason we study such evolution for the three cases, namely, weak coupling, strong
coupling and resonance case according to
$ (\lambda/\omega)<<1,
 (\lambda/\omega)>>1$ and $ (\lambda/\omega)=1$, respectively.

\section{Evolution of the squeezed regions of the PCS}
The PCS is a special type of squeezed vacuum state
\cite{eber} which is essentially equivalent to the two-photon coherent state
\cite{yu2}, and it possesses most of the properties of the ordinary
coherent states, such as a completeness relation and a reproducing kernel.
Also the PCS can be realized in the framework of degenerate and
nondegenerate parametric amplifier \cite{cher4}.
The required quantities for discussing the behaviour of $F_{x,y}(.)$ in
(\ref{i4}) are given by
\begin{eqnarray}
\begin{array}{lr}
 \langle (\Delta \hat{k}_{x}(t))^{2}\rangle = 2k \left\{
|f(t)|^{2}+\frac{[S(t)-\xi^{*}f(t)-\xi
f^{*}(t)]^{2}}{(1-|\xi|^{2})^{2}}
+\frac{S(t)[\xi^{*}f(t)+\xi f^{*}(t)-S(t)]}
{(1-|\xi|^{2})}\right\},
\\\\
 \langle (\Delta \hat{k}_{y}(t))^{2}\rangle = 2k \left\{
|G(t)|^{2}+\frac{[V(t)+\xi^{*}G(t)+\xi
G^{*}(t)]^{2}}{(1-|\xi|^{2})^{2}}
-\frac{V(t)[\xi^{*}G(t)+\xi G^{*}(t)+V(t)]}
{(1-|\xi|^{2})}\right\}, \\\\
\langle \hat{k}_{z}(t)\rangle = \frac{k}{(1- |\xi|^{2})} \left\{
(1+ |\xi|^{2})
R_{3}(t) + 2[\xi^{*}h(t)+\xi h^{*}(t)]\right\},
\label{ev1}
\end{array}
\end{eqnarray}
 where

\begin{equation}
f(t)=\frac{1}{2} [ R_{1}(t)-iJ(t)], \quad G(t)=\frac{1}{2} [J(t)
-iR_{2}(t)],\quad h(t)=\frac{1}{2}[S(t) -iV(t)].\label{ev2}
\end{equation}
From (\ref{ev1}) and (\ref{i4}) it is evident that the fluctuations are
independent of
the Bargmann index $k$.

Initially,  at $t=0$, the results  (\ref{ev1}), (\ref{ev2}) with
  (\ref{i4}) lead to the following expressions:

\begin{equation}
F_{x}(r,\Phi,t=0)=\frac{\tanh^{2}(\frac{r}{2})}
{1-\tanh^{4}(\frac{r}{2})}\left[(1+\cosh r
)\cos^{2}\Phi-1\right],
\label{1a}
\end{equation}
and
\begin{equation}
F_{y}(r,\Phi,t=0)=\frac{\tanh^{2}(\frac{r}{2})}
{1-\tanh^{4}(\frac{r}{2})}\left[(1+\cosh r
)\sin^{2}\Phi-1\right].
\label{1b}
\end{equation}
The condition for squeezing the $\hat{k}_{x}$- or the
$\hat{k}_{y}$-component is $F_{x}<1$ or $F_{y}<1$, namely, the results
first obtained in \cite{eber},

\begin{equation}
1+\cos^{2}\Phi \sinh^{2}(2r)\leq \cosh (2r)
\label{1ab}
\end{equation}
or
\begin{equation}
1+\sin^{2}\Phi \sinh^{2}(2r)\leq \cosh (2r),
\label{1bb}
\end{equation}
respectively.

It is clear from (\ref{1a}) and (\ref{1b}) that the squeezed regions
 are symmetric in $r$
 and  periodic in
$\Phi$ with period $\pi$ , i.e.,
$F_{j}(r,\Phi,t=0)=F_{j}(-r,\Phi+\pi,t=0)$.
Furthermore,  $F_{x}(.)$ connects with $F_{y}(.)$ through the relation

\begin{equation}
F_{y}(r,\Phi,t=0)=
F_{x}(r,\Phi+\frac{\pi}{2},t=0)
\label{1c}
\end{equation}
and thus
have similar behaviour with $F_{y}$
 shifted by $\pi/2$ along the $\Phi$-axis.
\begin{figure}
{\includegraphics[width=8cm]{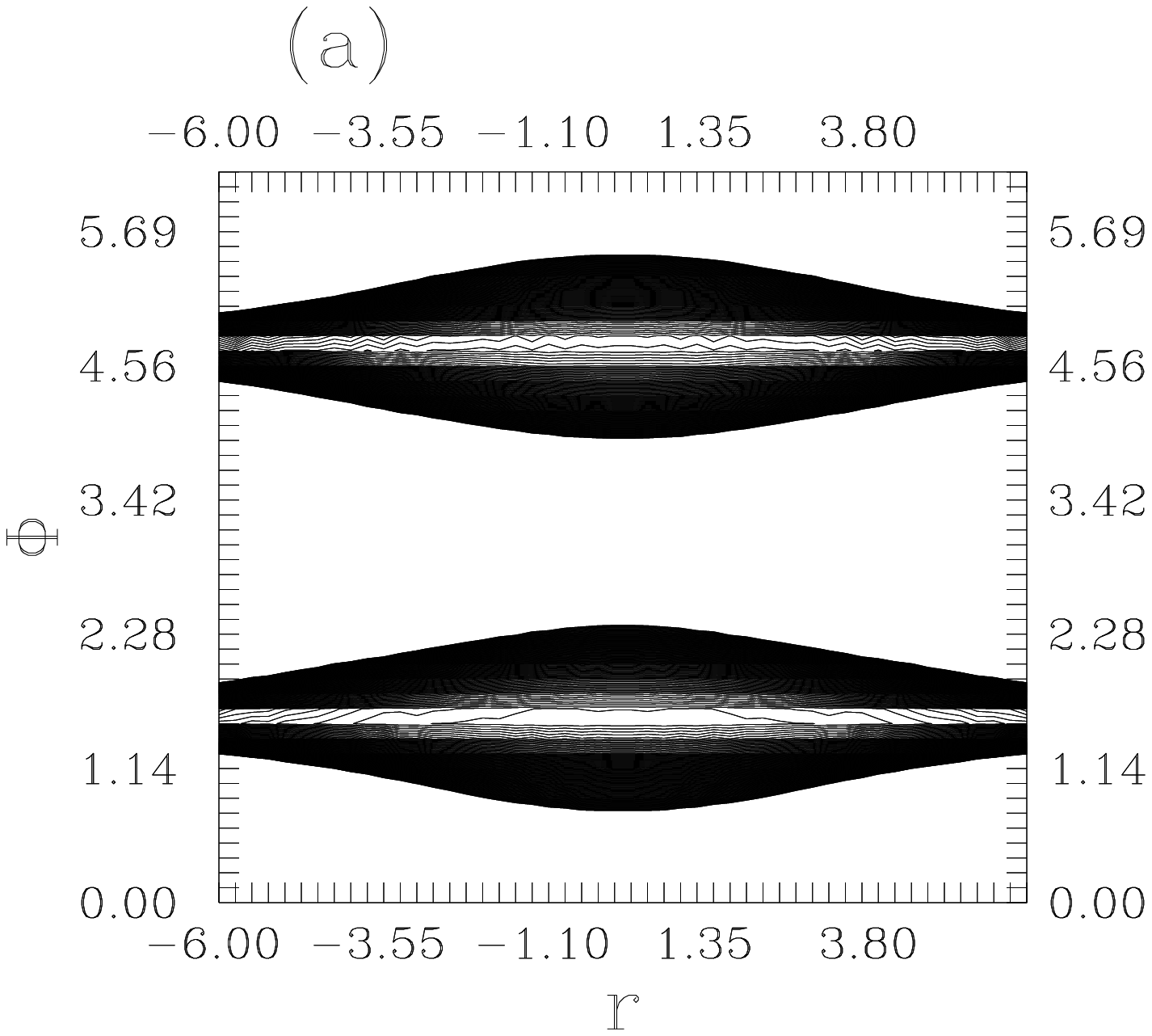}}
{\includegraphics[width=8cm]{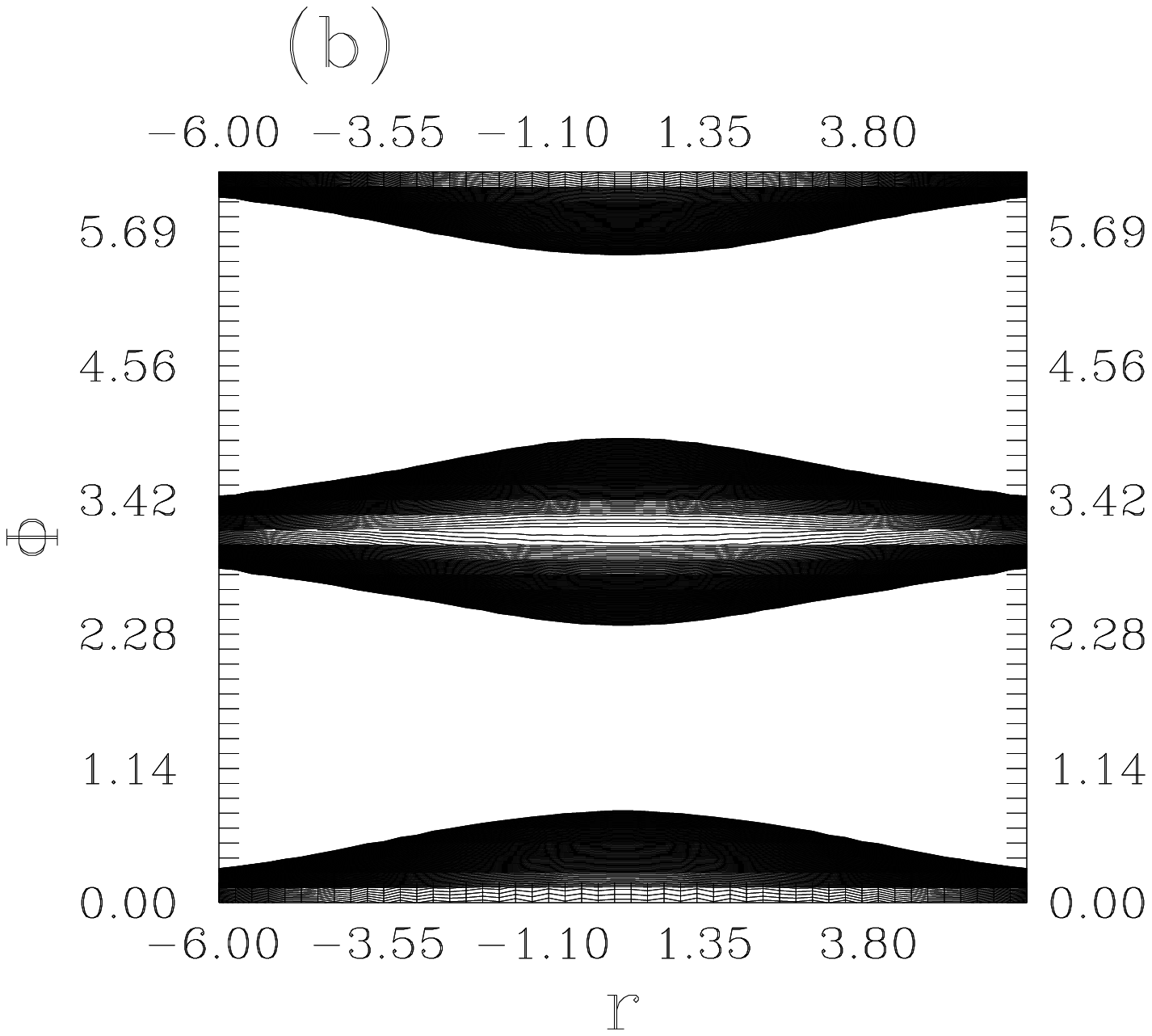}}
 \caption{ The squeezing
regions in the $(r,\Phi)$-plane of the $\hat{k}_{x}$- and
$\hat{k}_{y}$-components  $(a)$ and $(b)$, respectively, at
initial time $t=0$.}
\end{figure}
It is also clear  from (\ref{1a}) that  $F_{x}(.)$ has maximum squeezing
($100 \%$ )  at $\Phi=\pi/2$ and $3\pi/2$ regardless of the values of
$r$ and it is symmetric around these lines, i.e.,
$F_{x}(.,\Phi=\pi/2+\epsilon,t=0)=F_{x}(.,\Phi=\pi/2-\epsilon, t=0)$ where
$\epsilon\leq \pi/2$ \cite{eber}.  Furthermore, squeezed intervals
on the specific lines can be obtained  by applying
the squeezing conditions $F_{j}\leq 0$, e.g.  setting $r=0$ in
(\ref{1a}) and analyzing the squeezing inequality lead to

\begin{equation}
\frac{3\pi}{4}\geq \Phi\geq \frac{\pi}{4}, \qquad
\frac{7\pi}{4}\geq \Phi\geq \frac{5\pi}{4}. \label{2a}
\end{equation}
Restricting our discussion to the first region (where both of which are
similar) we see that at the borders of this region the state is  minimum-uncertainty state
 but when $\Phi$ "evolves" the squeezing values increase gradually
 showing its maximum
value at  $\Phi=\pi/2$, i.e. maximum squeezing occurs at the most inner contours.
All this information  is clear in Figs. 1 where we
have plotted the squeezed regions for $F_{x}(.)$ and $F_{y}(.)$ in
the $(r,\Phi)$-plane for given values of the parameters.
Comparison of Fig. 1a and 1b leads to the fact that there are specific regions in the
$(r,\Phi)$-plane for which neither the $\hat{k}_{x}$- nor
the $\hat{k}_{y}$-quadrature is squeezed. Furthermore,  squeezing
cannot be established in the two quadratures simultaneously.

Now for $t>0$ we discuss the evolution of squeezed regions
of $F_{j}(r,\Phi,t)$ for the  system under consideration in
$(r,\Phi)$-plane for the three cases mentioned above.
\begin{figure}
  {\includegraphics[width=8cm]{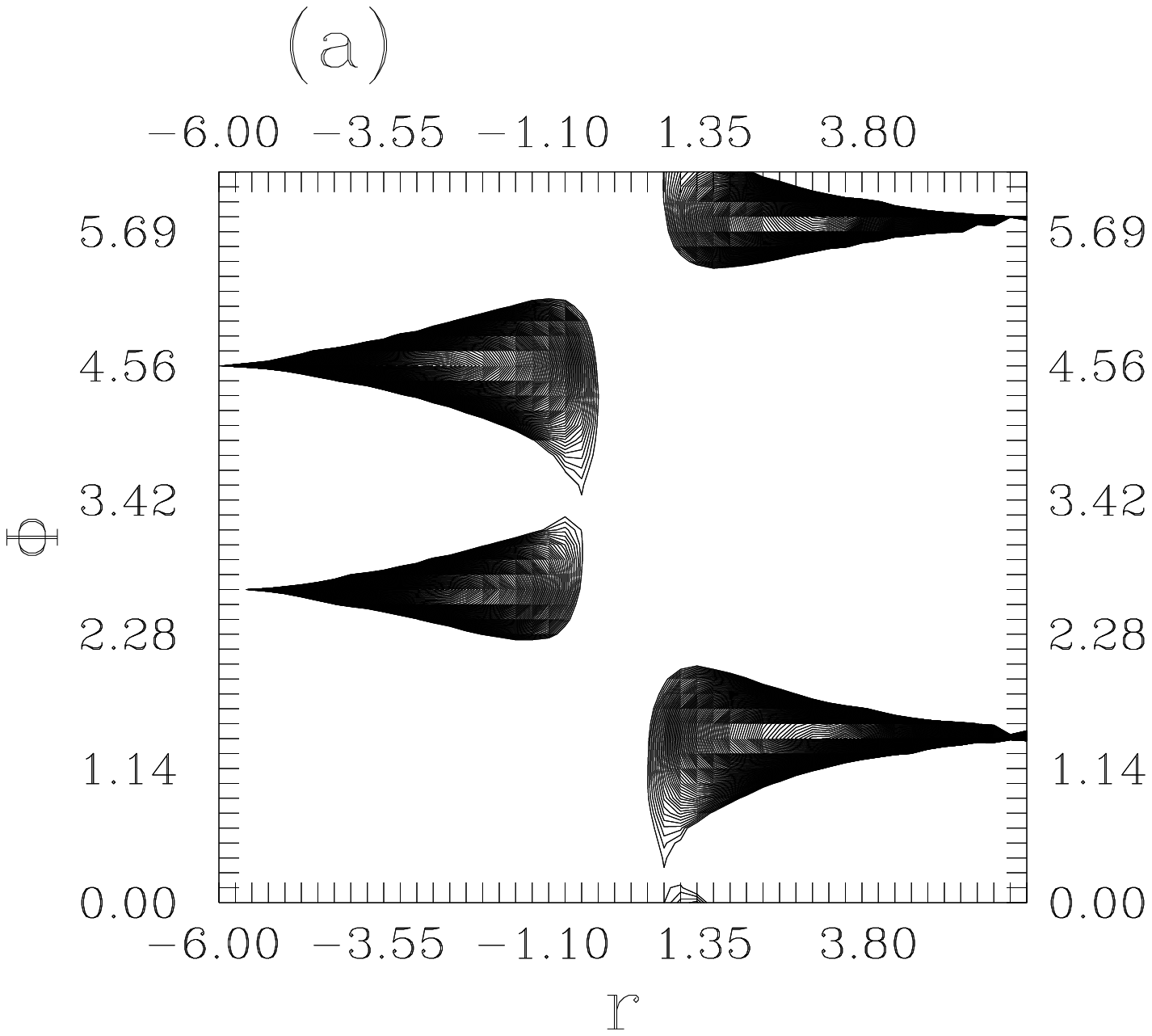}}
  {\includegraphics[width=8cm]{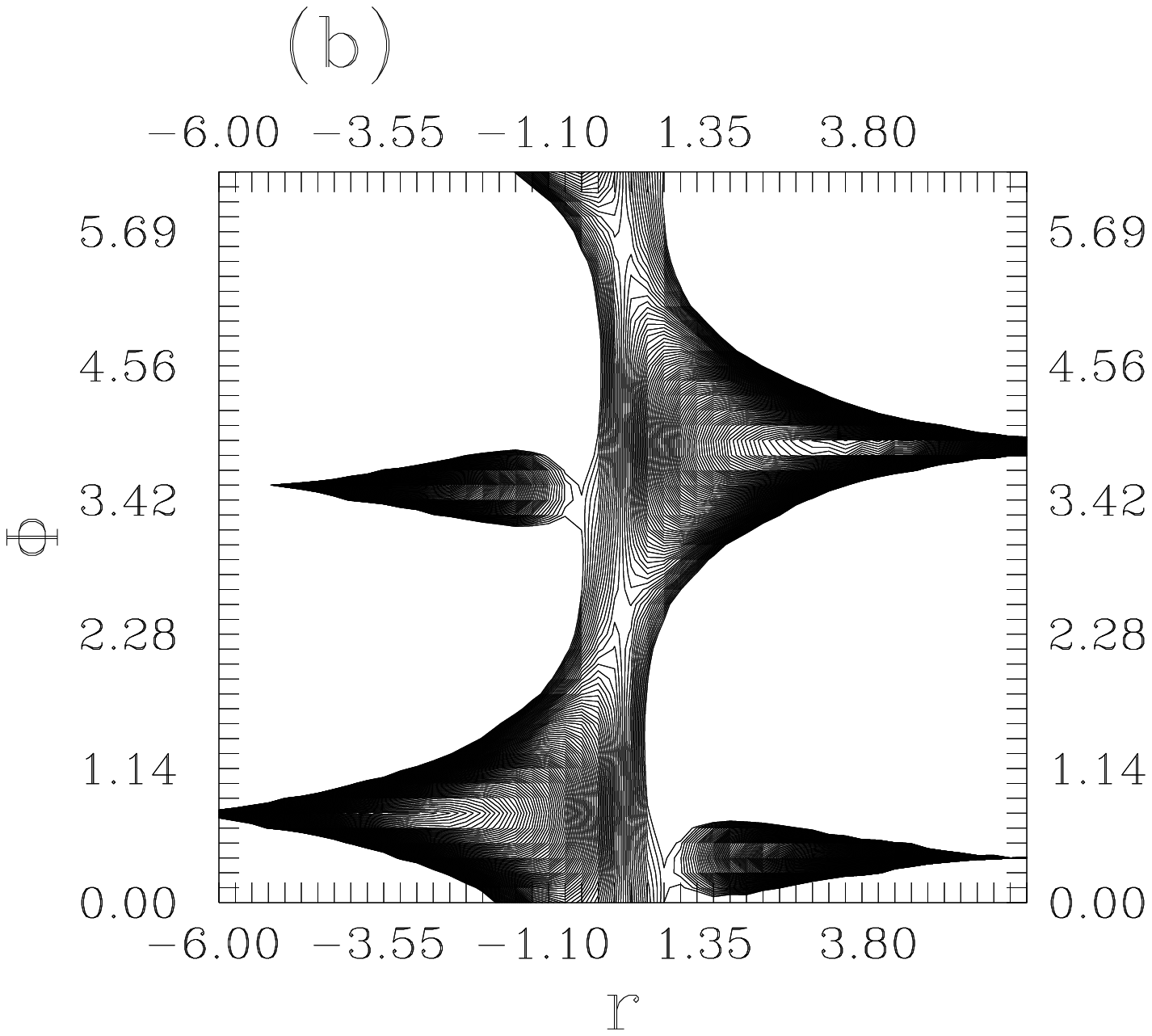}}
 \caption{ The squeezing regions in the $(r,\Phi)$-plane of the
$\hat{k}_{x}$- and $\hat{k}_{y}$-components  $(a)$ and $(b)$,
respectively,  for the normalized time
$t_{\lambda}=t\lambda=\pi/2$ and $\omega/\lambda=3$}
\end{figure}
\subsection{weak coupling}
In this case $ (\lambda/\omega)<<1$, i.e. the frequency detuning  parameter
$\omega$ is much greater than the coupling constant $\lambda$.
The exact expressions for the variances  in
(\ref{ev1}) can be obtained
easily for the zeroth order of $ (\lambda/\omega)$, which have similar
forms as those of the initial ones (\ref{1a}) and (\ref{1b}) but with
$\Phi\rightarrow \Phi+2\omega t$. This means that under this condition
the initial state $|\xi,k\rangle$ evolves into the dynamical state (up
to a constant phase) $|\xi\exp(2i\omega t),k\rangle$. More
illustratively,
the free part in the Hamiltonian (\ref{f1}) acts much effectively
 on the behaviour of the dynamical system  than the
interaction part and then the dynamical state of the system can be expressed as:
\begin{equation}
\hat{U}(t)|\xi,k\rangle
\simeq\exp(2i\omega t\hat{k}_{z})|\xi,k\rangle
=|\xi\exp(2i\omega t),k\rangle. \label{fw1}
\end{equation}
So it is obvious that in this case the initial squeezed
regions move along the $\Phi$-axis by an ammount $2\omega t$ having typical
shapes and sizes as those of the initial regions.
This situation is similar to that of the $SU(2)$ systems \cite{hassan1}
in atomic coherent state.
For completeness,
we provide  Figs. 2a,b and Figs. 3a,b for the squeezed regions of the exact forms
(\ref{ev1}) when $(\omega/\lambda)=3$ and $10$, respectively.
From Figs. 2 it seems that  the sizes of
the squeezed regions are decreased compared to those of the initial ones and
also the  symmetry in the $(r,\Phi)$-plane is smeared out.
These facts are remarkable by comparing Figs. 1 and 2.
On the other hand,
by increasing the values of $\omega$ compared to those of $\lambda$
the behaviour of the system starts to going to that of the initial
states (compare Figs. 1 and 3) and this agrees with the discussion
  given above (c.f. (\ref{fw1})).
Furthermore,
generally for  the weak coupling case
 the initial squeezing is periodically restored and
 the maximum squeezing occurs at the most inner contours.

\begin{figure}
  {\includegraphics[width=8cm]{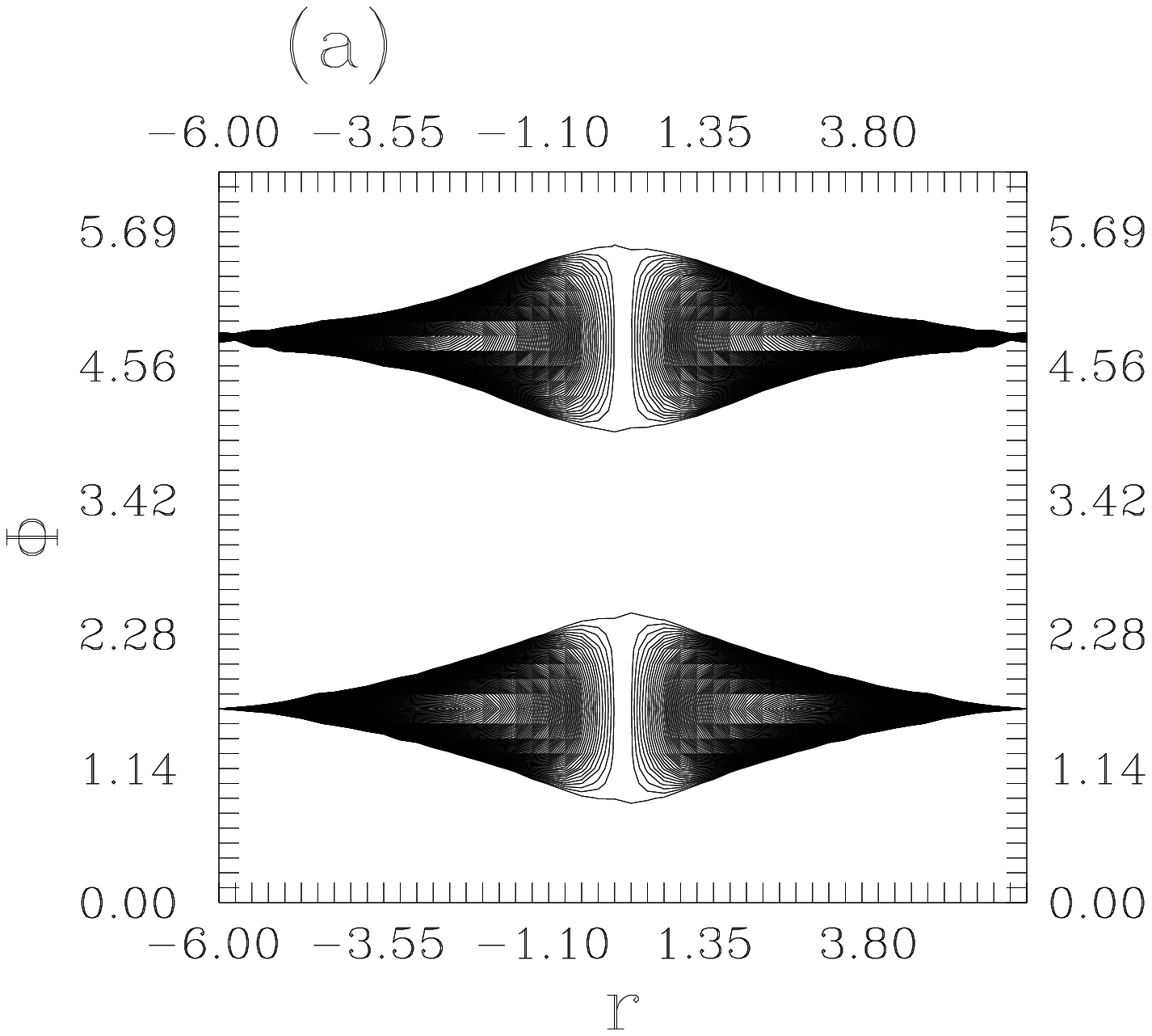}}
  {\includegraphics[width=8cm]{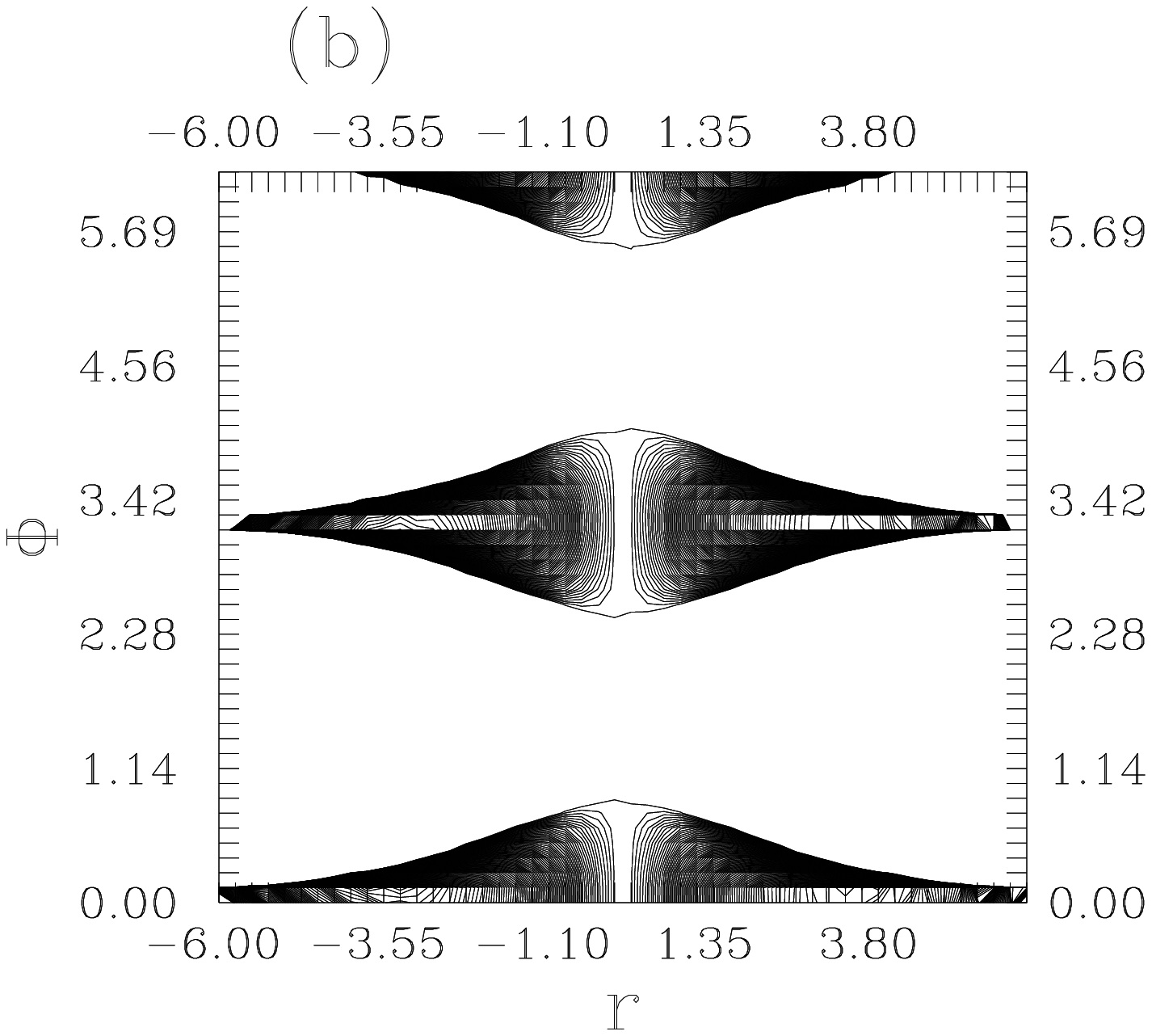}}
 \caption{ The squeezing regions in the $(r,\Phi)$-plane of the
$\hat{k}_{x}$- and $\hat{k}_{y}$-components  $(a)$ and $(b)$,
respectively,  for the normalized time
$t_{\lambda}=t\lambda=\pi/2$ and $\omega/\lambda=10$}
\end{figure}

\subsection{Strong coupling}
In this case $ (\lambda/\omega)>>1$, i.e.
the coupling constant $\lambda$ is much greater than
the detuning  parameter $\omega$.
In this case the Rabi frequency parameter $g$ tends to
$i\sqrt{\lambda^{2}-\omega^{2}}$ and consequently the trigonometric functions
in the dynamical solutions (\ref{i6}) are converted into hyperbolic functions,
which are
monotonically increasing functions of $t$ and then the initial squeezing
of the PCS will  vanish even for very short interaction time.
For instance, for large interaction time $t$ (taking normalized time
$\tau=gt$),  $\cosh(\tau)\simeq\sinh(\tau)\simeq \frac{1}{2}\exp(\tau)$ and
(\ref{ev1}) reduce to such types of proportionality
\begin{equation}
 \langle (\Delta \hat{k}_{j}(t))^{2}\rangle \propto \exp(4\tau),\quad j=x,y,\quad
  \langle \hat{k}_{z}(t)\rangle\propto\exp(2\tau). \label{fww1}
\end{equation}
This shows that the initial squeezed quadrature ($F_{j}(t=0)<0$)
becomes unsqueezed due to the interaction with the material media
regardless of the values of $\xi$.
This situation is completely
in contrast with the $SU(2)$ case \cite{hassan1} where squeezing always exists
and the initial squeezing values are periodically recovered.
It is a feature difference between  the  $SU(2)$ and
$SU(1,1)$ Lie algebraic structures.
\begin{figure}
  {\includegraphics[width=8cm]{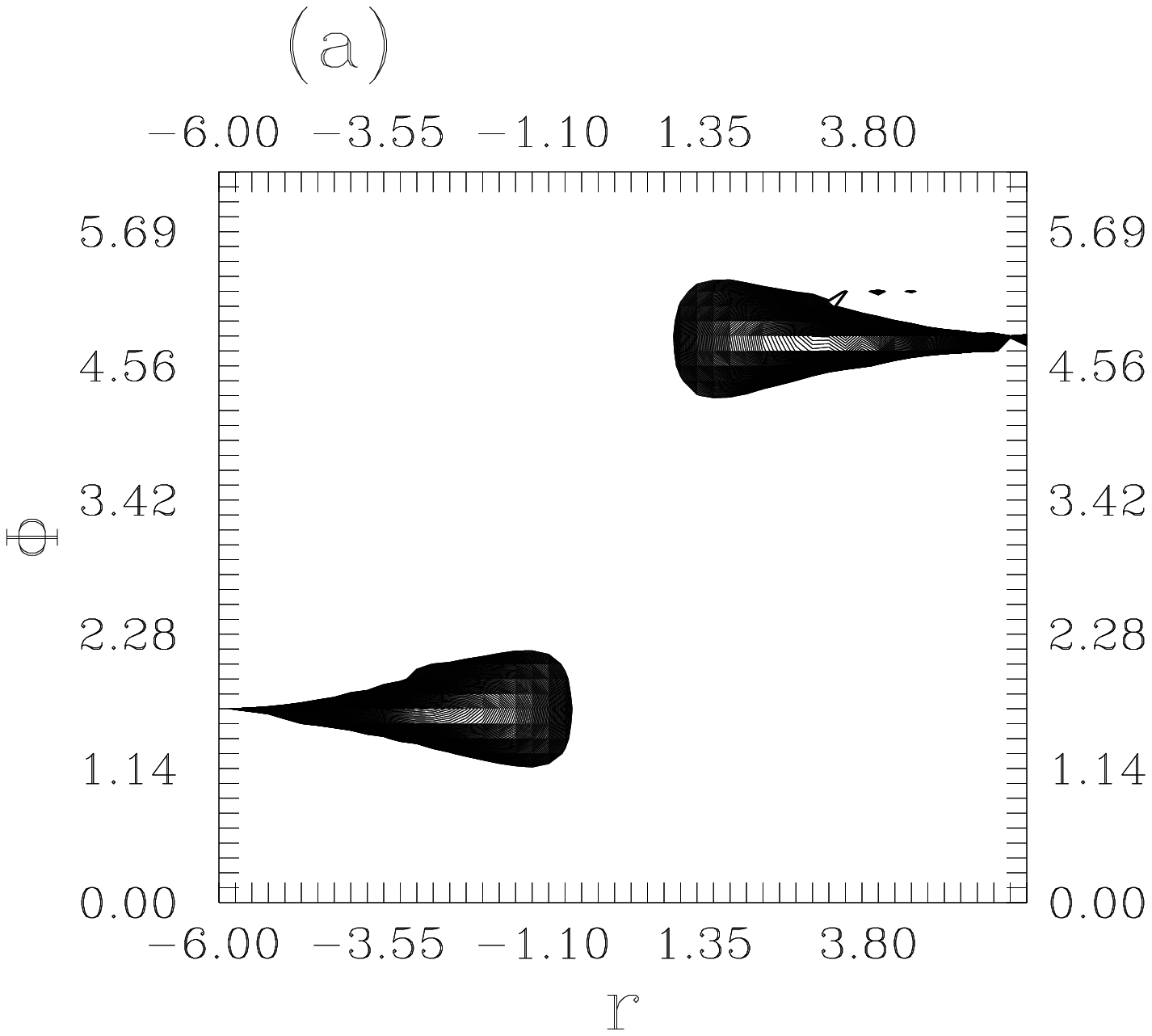}}
  {\includegraphics[width=8cm]{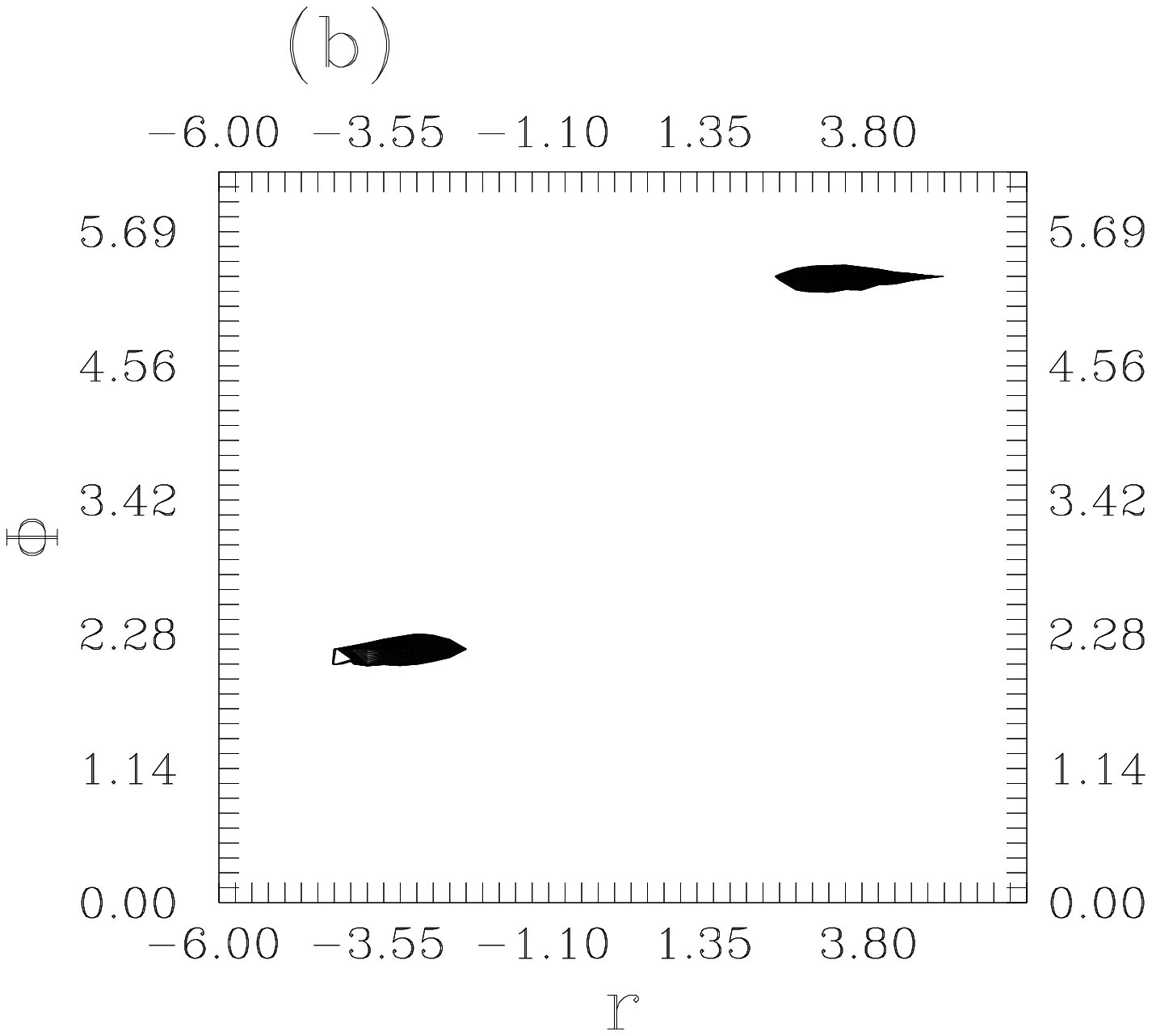}}
   \caption{ The
squeezed regions in the $(r,\Phi)$-plane of the $\hat{k}_{x}$- and
$\hat{k}_{y}$-components  $(a)$ and $(b)$, respectively, at
$t_{\omega}=t\omega=\pi/4$ and $\lambda/\omega=2$ .}
\end{figure}
In  Figs. 4a and b  we have plotted the squeezed regions for
 the $\hat{k}_{x}$- and
$\hat{k}_{y}$-components.
In these figures we have considered  $\lambda$ is greater than $\omega$
 but not too much. Comparison of Figs. 1 and 4 leads to
the fact that  the sizes of
the squeezed regions are decreased compared with those of the initial ones.
Furthermore, the squeezed area regarding to the $\hat{k}_{y}$-component is
decreased more rapidly than that in the
$\hat{k}_{x}$-component. This is a direct consequence of the structure of
the Hamiltonian (\ref{f1}). Also the
symmetry in the $(r,\Phi)$-plane for the chosen values
is  remarkable.
\subsection{Resonance  case}
In this case $ (\lambda/\omega)=1$, i.e.
the coupling constant $\lambda$ is equal to
the detuning  parameter $\omega$. This manifests itself in the quadrature
variances, which become polynomials of the normalized time $\tau=t\lambda=t\omega$.
In this case expressions (\ref{ev1}) reduce to
\begin{eqnarray}
\begin{array}{lr} \langle (\triangle
\hat{k}_{x}(\tau))^{2}\rangle =2k\Bigl\{ \tau^{4} \chi^{2}
+\tau^{2}[\chi\sinh r \cos\Phi+\sinh^{2}r \sin^{2}\Phi]
\\
\\
-2\tau^{3}\chi \sinh r \sin\Phi
-\frac{\tau}{2}\sinh^{2}r \sin(2\Phi)+\varepsilon(\Phi)\Bigr\}
, \label{rc1}
\end{array}
\end{eqnarray}

\begin{equation}
\langle (\triangle \hat{k}_{y}(\tau))^{2}\rangle
=2k\Bigl\{ \tau^{2} \chi^{2}
-\tau\chi\sinh r \sin\Phi
+\varepsilon(\Phi+\frac{\pi}{2})\Bigr\},
 \label{arc2}
\end{equation}
\begin{equation}
\langle  \hat{k}_{z}(\tau)\rangle
=2k\Bigl\{ \tau^{2} \chi
-\tau\sinh r \sin\Phi
+\frac{1}{2}\cosh r\Bigr\},
 \label{arc3}
\end{equation}
where
\begin{equation}
 \chi=\cosh r
-\sinh r \cos\Phi,\quad
\varepsilon(\Phi)=\frac{1}{4}[1+\sinh^{2}r \cos^{2}\Phi].
 \label{arc4}
\end{equation}
It is evident that
 (\ref{rc1}) is a polynomial of order four in $\tau$, however,
(\ref{arc2}) and (\ref{arc3})  are polynomials of order two. This leads to
(one can easily check) when $\tau>>1$, $F_{y}(.)$ exhibits always squeezing,
 which becomes time-independent (steady-state squeezing).
 Nevertheless, this is not the case for $F_{x}(.)$
where squeezing is completely suppressed as the time evolves.
For $\tau>>1$ one can deduce  straightforwardly
that $F_{y}(.)$ takes the form
\begin{equation}
F_{y}(r,\Phi)=2\chi-1. \label{arc5}
\end{equation}
In other words,  the squeezing condition for the $\hat{k}_{y}$-component is
\begin{equation}
2(\cosh r
-\sinh r \cos\Phi)-1\leq 0. \label{arc6}
\end{equation}
Inequality (\ref{arc6}) fails for
 $\Phi=m\pi/2,\quad m=1,3,5,..$ regardless  the value of $r$ and also
 for $r=0$ regardless  the value of $\Phi$.
On the contrary, (\ref{arc6}) is satisfied
for both $r>0$ and $\Phi=m\pi,\quad m=0,2,4,..$ and also for
 $r<0$ and $\Phi=m\pi,\quad m=1,3,5,..$.
 Inequality
(\ref{arc6}) can be solved easily to some representative lines.
For example, on $\Phi=0$ (or  $\Phi=\pi$)
it
 gives $r\succeq 0.69$, where
 $F_{y}(.,r\simeq 0.69,.)\simeq 0$.
Furthermore,  maximum value
of  squeezing can be obtained when $F_{y}=-1$, i.e. when
$\chi=0$ and this leads to
$ \cos\Phi =\coth r$. That is when $\Phi=0$,
$r\rightarrow \infty$  maximum squeezing is achieved.
Summing up  on the line $\Phi=0$ the system exhibits minimum-uncertainty
state at $r\simeq 0.69$ and then the values of the
 squeezing gradually increase as $r$ increases.
 All these analytical facts are noticeable in
 Fig. 5, where we have plotted $F_{y}(.)$.  On the other hand, as we
 have mentioned above, the initial
 squeezing inherited in $F_{x}(.)$ is suppressed gradually as the
 interaction evolves. For example,
on the line $r=0$ squeezing exists only when $\tau\leq 1/\sqrt{2}$
and also maximum squeezing cannot be occurred.

\begin{figure}
  {\includegraphics[width=8cm]{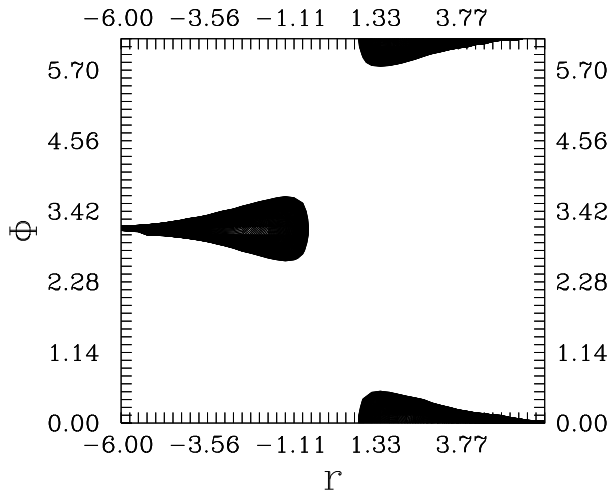}}
\caption{ The squeezed regions in the $(r,\Phi)$-plane of the
$\hat{k}_{y}$-component for  normalized time $\tau=3\pi$ in the
resonance case.}
\end{figure}
\section{Evolution of the squeezed regions of the BGCS}
In this section we investigate the evolution of the squeezed regions in the
plane $(|Z|,\Phi)$  when the  system is initially in the state BGCS given by
(\ref{i8}).
It is worth mentioning that the BGCS has  similar properties as
the ordinary (Glauber) coherent state in the sense
that it is  unsqueezed
state and  a minimum-uncertainty state (so-called $SU(1,1)$ intelligent state).
The required quantities appearing in the expressions of
$F_{x,y}(t)$ in (\ref{i4}) for the BGCS are given by

\begin{eqnarray}
\begin{array}{lr}
\langle (\Delta \hat{k}_{x}(t))^{2}\rangle =
2|f(t)|^{2} \left[k+ \frac{|Z| I_{2k}(2|Z|)}{I_{2k-1}(2|Z|)}\right]-
S(t)[Z^{*}f(t)+Z f^{*}(t)]
 \\\\
 + |Z| S^{2}(t)\left[ |Z|\left(1-
\frac{ I^{2}_{2k}(2|Z|)}{I^{2}_{2k-1}(2|Z|)}\right)+(1-2k)
\frac{ I_{2k}(2|Z|)}{I_{2k-1}(2|Z|)}\right],
\\\\
 \langle (\Delta \hat{k}_{y}(t))^{2}\rangle =
2|G(t)|^{2}\left[k+ \frac{|Z| I_{2k}(2|Z|)}{I_{2k-1}(2|Z|)}\right]-
V(t)[Z^{*} G(t)+Z G^{*}(t)]
\\\\
+ |Z| V^{2}(t)\left[ |Z|\left(1-
\frac{ I^{2}_{2k}(2|Z|)}{I^{2}_{2k-1}(2|Z|)}\right)+(1-2k)
\frac{ I_{2k}(2|Z|)}{I_{2k-1}(2|Z|)}\right],
\\\\
 \langle \hat{k}_{z}(t)\rangle =
 R_{3}(t) \left[k+ \frac{|Z| I_{2k}(2|Z|)}{I_{2k-1}(2|Z|)}\right]
+ Z^{*}h(t)+Z h^{*}(t), \label{sct1}
\end{array}
\end{eqnarray}
 where $f(t), G(t)$ and $h(t)$ are given in
(\ref{ev2}).
Initially ($t=0$) from (\ref{sct1}) one recovers the no-squeezing
minimum uncertainty relation $\langle (\Delta \hat{k}_{x,y}(0))^{2}\rangle=
\frac{1}{2}|\langle \hat{k}_{z}(0)\rangle|$ derived in \cite{(19)}. For $t>0$
and throughout  the discussion of the present section
 we emphasize on the two limiting cases, namely,
 the weak and strong intensity limits,  $|Z|<<1$ and
$|Z|>>1$, respectively, provided that the interaction time and
 the Bargmann index $k$
are finite. These two approaches will give a good visualization for the
behaviour of the system in a whole $(|Z|,\Phi)$-plane.
For the weak intensity case (\ref{sct1}) can be simplified into the forms
\begin{eqnarray}
\begin{array}{lr} \langle (\Delta \hat{k}_{x}(t))^{2}\rangle
\simeq [2|f(t)|^{2}+S^{2}(t)] \frac{|Z|^{2}}{2k} +2k|f(t)|^{2}-
S(t)[Z^{*}f(t)+Z f^{*}(t)],
\\\\
 \langle (\Delta \hat{k}_{y}(t))^{2}\rangle \simeq
[2|G(t)|^{2}+V^{2}(t)] \frac{|Z|^{2}}{2k} +2k|G(t)|^{2}-
V(t)[Z^{*}G(t)+Z G^{*}(t)],
\\\\
 \langle \hat{k}_{z}(t)\rangle \simeq
 R_{3}(t) (k+ \frac{|Z|^{2}}{2k})
+ Z^{*}h(t)+Z h^{*}(t), \label{sct2}
\end{array}
\end{eqnarray}
while for the strong intensity case they take the forms
\begin{eqnarray}
\begin{array}{lr} \langle (\Delta \hat{k}_{x}(t))^{2}\rangle
\simeq 2|f(t)|^{2}(k+ |Z|)- S(t)[Z^{*}f(t)+Z f^{*}(t)]
 + |Z|(1-2k) S^{2}(t),
\\\\
 \langle (\Delta \hat{k}_{y}(t))^{2}\rangle \simeq
2|G(t)|^{2}(k+ |Z|)-
V(t)[Z^{*} G(t)+Z G^{*}(t)]
+ |Z|(1-2k) V^{2}(t),
\\\\
 \langle \hat{k}_{z}(t)\rangle \simeq
 R_{3}(t) (k+ |Z| )
+ Z^{*}h(t)+Z h^{*}(t). \label{sct3}
\end{array}
\end{eqnarray}
\begin{figure}
  {\includegraphics[width=8cm]{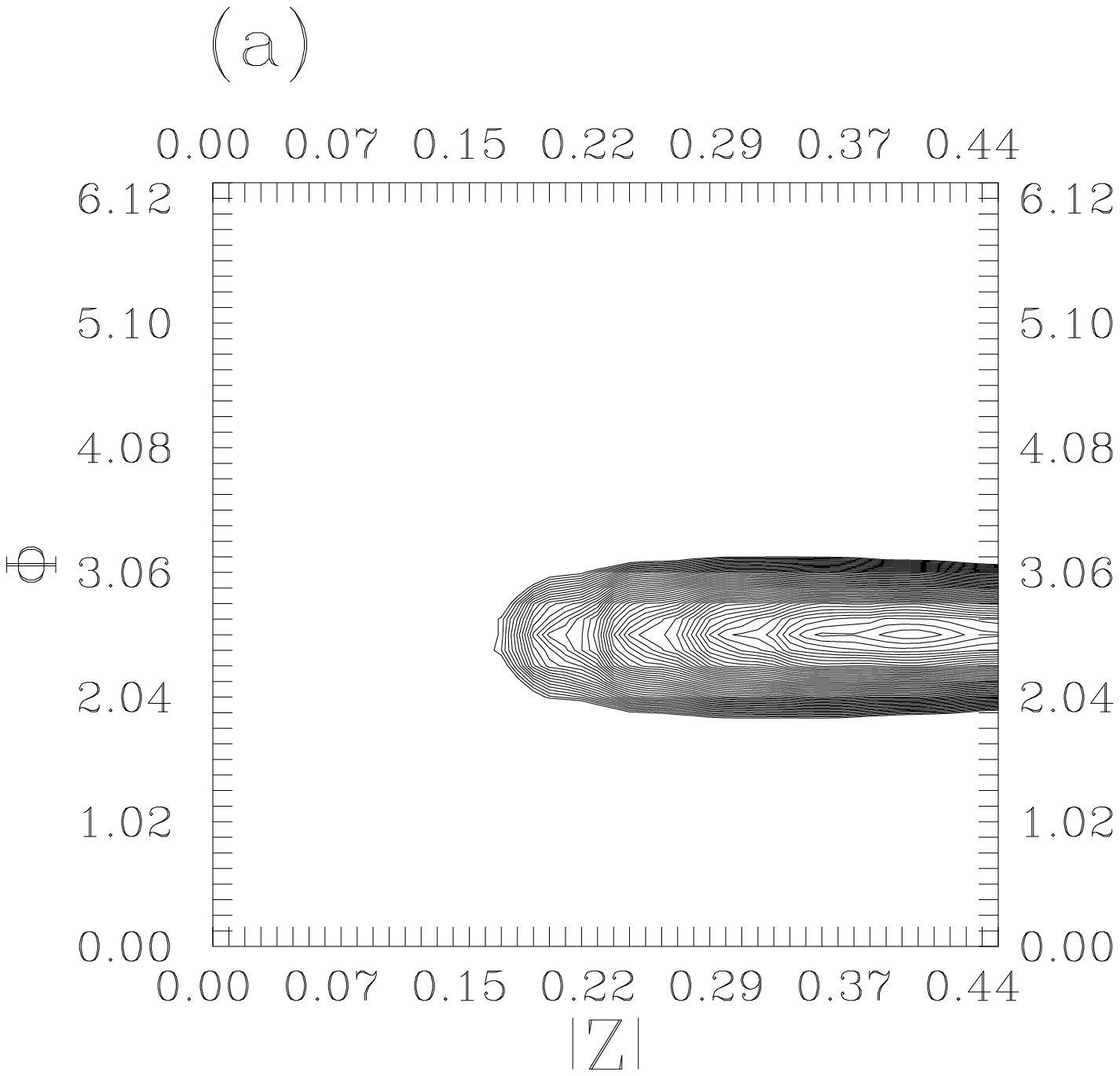}}
  {\includegraphics[width=8cm]{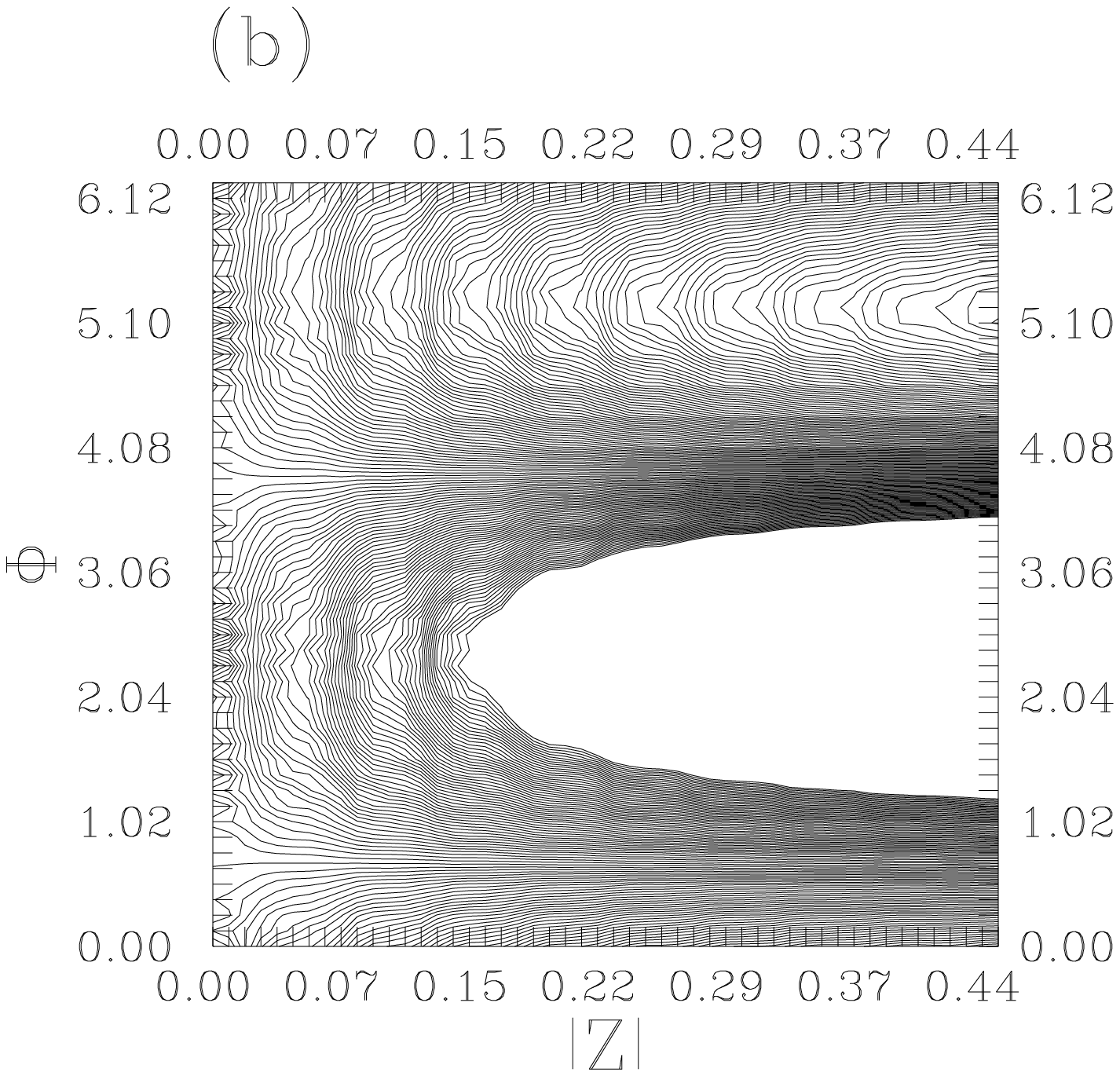}}
 \caption{ The squeezed regions in the $(|Z|,\Phi)$-plane of the
$\hat{k}_{x}$- and $\hat{k}_{y}$-components   $(a)$ and $(b)$,
respectively, for $k=0.5,t_{\lambda}=t\lambda=1$ and
$\omega/\lambda=3$.}
\end{figure}
In the following  we  discuss the different cases as we did in section 3.
\subsection{weak coupling}
As we mentioned  above the BGCS is not a squeezed state. So that following similar
discussion as that given in subsection 3.1 one can prove  that for
weak coupling case the
system evolves always in a minimum-uncertainty state.
Now we discuss the case when $\omega$ is greater than $\lambda$ but not
too much. In this case we have noted that squeezing  can be occurred.
This of course  arises from the interaction of the field  with the
nonlinear medium.
This situation can be recognized analytically
when $|Z|\rightarrow 0$, the quantities  $F_{j}(.)$
reduce to
\begin{eqnarray}
\begin{array}{lr}
F_{x}(\tau)=\frac{2\lambda^{2}}{g^{2}R_{3}(\tau)}\left[
\frac{2\omega^{2}}{g^{2}}
\sin^{2}(\tau)-1\right]\sin^{2}(\tau),
\\\\
F_{y}(\tau)=\frac{2\lambda^{2}}{g^{2}R_{3}(\tau)}
\cos(2\tau)\sin^{2}(\tau), \label{sct4}
\end{array}
\end{eqnarray}
where $\tau=gt$.  We have to stress that $R_{3}(t)$
is always positive (c.f. (\ref{i6})). From (\ref{sct4})  squeezing can
occur in  $F_{x}$ or $F_{y}$, at particular values of the
interaction parameters, e.g. maximum squeezing can be generated in $F_{y}$
 at $\tau=\pi/2$. Also when $(\lambda/\omega)<<1$,
 (\ref{sct4}) give   $F_{x,y}\simeq
0$, which agree with the above conclusion.
Information about the weak intensity case
is shown in Figs. 6 for given values of the system parameters.
 Maximum squeezing in the $\hat{k}_{x}$-component occurs  at the
most inner contour, however,
for the $\hat{k}_{y}$-component it occurs around the line
$\Phi\simeq 7\pi/4$.
Also there is  a lack of symmetry in the phase plane somewhat similar
to the PCS case.
Generally, squeezing is sensitive to the values of Bargmann index $k$,
in contrast with the PCS case, which decreases as $k$ increases.
\begin{figure}
  {\includegraphics[width=8cm]{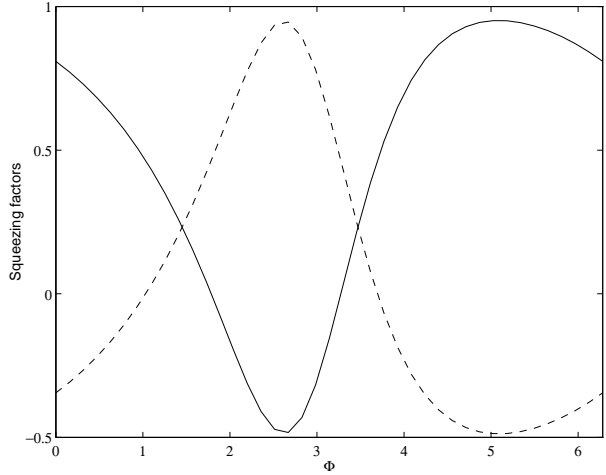}}
\caption{ Squeezing factors $F_{j}(.)$ against $\Phi$-axis the
$\hat{k}_{x}$-  and $\hat{k}_{y}$-components corresponding to
solid and dashed curves, respectively, for $|Z|=200,
k=0.5,t_{\lambda}=t\lambda=\pi/2$ and $\omega/\lambda=3$.}
\end{figure}
On the other hand, for strong intensity case and for  finite values of
$k$, $F_{j}(.)$ become $|Z|$-independent. This is obvious from
(\ref{sct3}) where all quantities are polynomials of  first order
in $|Z|$.
For this reason we study squeezing occurrence
regarding to $\Phi$ for fixed values of $|Z|$. This is given in Fig. 7
for the shown values of the parameters. From this figure one  observes
 that the maximum value of squeezing is $50\%$.
From figures 6 and 7 we see that the squeezed area of $F_{x}(F_{y})$
decreases (increases) as $|Z|$  gets larger and then it persists.

\subsection{Strong coupling}
In this case $\lambda>>\omega$ and the system has a time-dependent
amplifying nature, which decreases the amount  of squeezing (as we have seen
in the PCS case).
Nevertheless,  under certain conditions,
squeezing can be detected in the system. This can be seen analytically
in the limiting case $|Z|\rightarrow 0$, where $F_{j}(.)$ read
\begin{eqnarray}
\begin{array}{rl}
F_{x}(\tau)=\frac{2\lambda^{2}}{g^{2}R_{3}(\tau)}\left[
\frac{2\omega^{2}}{g^{2}}
\sinh^{2}(\tau)-1\right]\sinh^{2}(\tau),
\\\\
F_{y}(\tau)=\frac{2\lambda^{2}}{g^{2}R_{3}(\tau)}\cosh(2\tau)
\sinh^{2}(\tau), \label{sct5}
\end{array}
\end{eqnarray}
where $\tau=gt$.
From (\ref{sct5}) it is clear that  squeezing  occurs only in
the  $\hat{k}_{x}$-component, in particular, when the interaction time is small,
 specifically, for times such that,
\begin{equation}
\sinh(\tau)
\leq\frac{g}{\sqrt{2}\omega}. \label{sct6}
\end{equation}
\begin{figure}
  {\includegraphics[width=8cm]{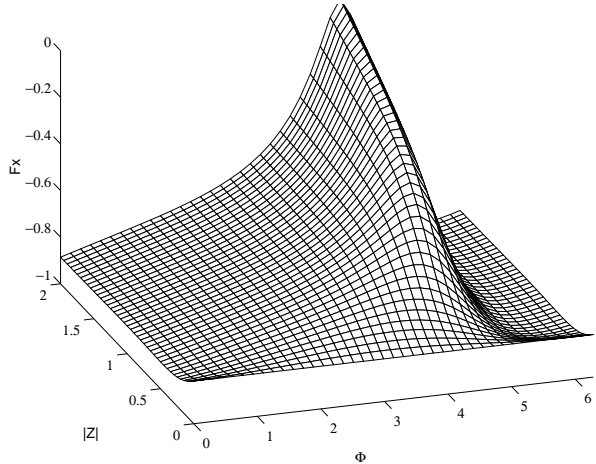}}\caption{
Squeezing factor $F_{x}(.)$  for $ k=0.5,t_{\omega}=
t\omega=\pi/20$ and $\lambda/\omega=10$.}
\end{figure}
In Fig. 8  we have plotted
$F_{x}(.)$ for given values of the interaction parameters.
Fig. 8 exhibits single-peak structure. More illustratively,
 maximum value of squeezing occurs
for $\Phi=0,2\pi$, however, on the line   $\Phi=\frac{3}{2}\pi$
squeezing values decrease monotonically as
 $|Z|$ increases,  eventually  suppressed
 and hence  $F_{x}(.)$ becomes
$|Z|$-independent.  Further, squeezing values are sensitive to the
Bargmann index $k$, in particular, for finite $|Z|$. For instance, as $k$
increases the squeezed region in $(|Z|,\Phi)$-plane increases, too.
\subsection{Resonance  case}
As in  section 3, at resonance ($\lambda=\omega$)  the variances and
the expectation values
are polynomials in the normalized time $\tau=t\lambda=t\omega$.
In the very weak intensity limit, $|Z|\rightarrow 0$, one obtains
\begin{figure}
  {\includegraphics[width=8cm]{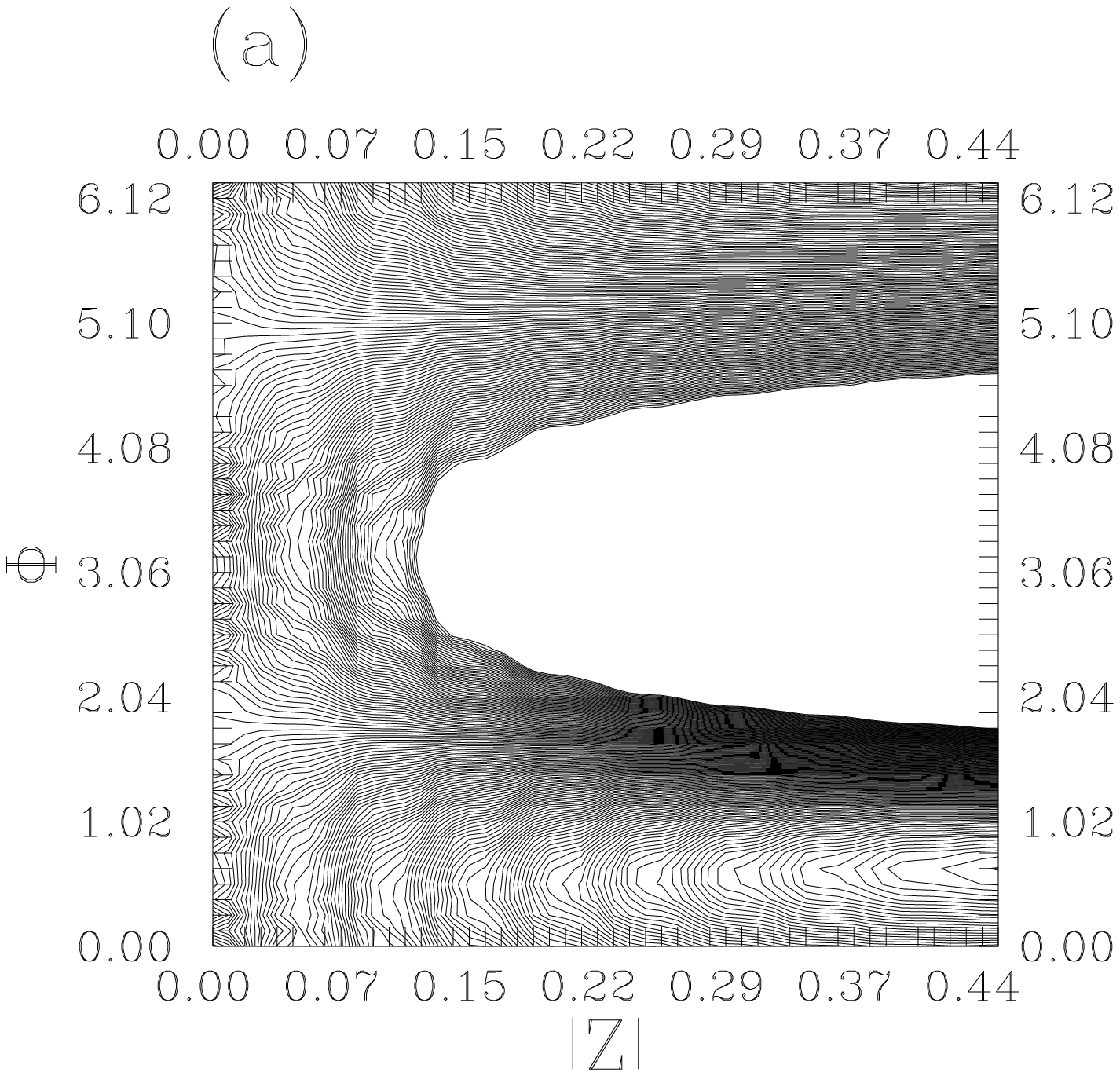}}
  {\includegraphics[width=8cm]{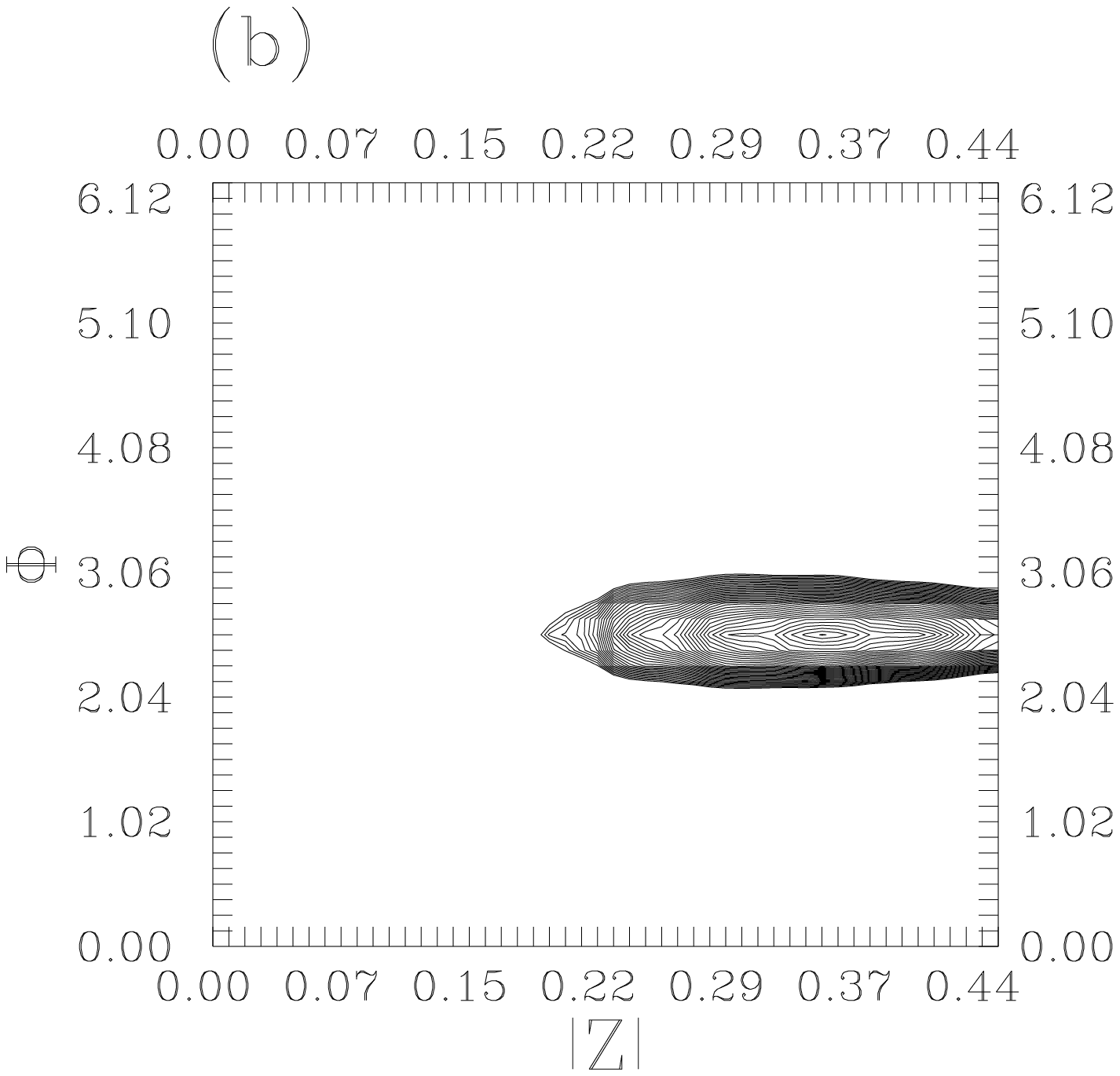}}
 \caption{ The
squeezed regions in the $(|Z|,\Phi)$-plane of the $\hat{k}_{x}$-
and $\hat{k}_{y}$-components   $(a)$ and $(b)$, respectively, for
$ k=0.5$ and $\tau=\pi/6$.}
\end{figure}
\begin{equation}
F_{x}(\tau)=\frac{2\tau^{2}(2\tau^{2}-1)}{(2\tau^{2}+1)}, \quad
F_{y}(\tau)=\frac{\tau^{2}}{(2\tau^{2}+1)},
 \label{sct6}
\end{equation}
which shows that
$F_{x}(.)$ exhibits squeezing for $\tau\leq 1/\sqrt{2}$.
On the other hand, numerical investigation shows that with the system
initially ($\tau=0$) in a non-squeezed
 minimum-uncertainty state and as the interaction is
switched  on
squeezing is  generated for a short interaction time in both quadratures components $\hat{k}_{x},
\hat{k}_{y}$ for weak intensity (, i.e.$|Z|<1$) (see Figs. 9)
and eventually  suppressed as the interaction time evolves.
Notice that the behaviour of $F_{x}(.)$ and $F_{y}(.)$ in the present
resonance case is the corresponding reverse behaviour in the weak
coupling case ($\omega >>\lambda$) of Fig. 6.
Similar behaviour has been noted for large values of $|Z|$ (see Fig. 7).
\section{Summary of the results}
Throughout this paper we have analysed the evolution of the squeezed regions
for the dynamical system whose Hamiltonian is a linear combination of the
$SU(1,1)$ Lie algebra generators. We have considered two initial states, namely,
the PCS and the BGCS. We have investigated the three cases: weak coupling
$(\omega/\lambda) >>1$, strong coupling $(\omega/\lambda) <<1$ and resonance
case $(\omega/\lambda) =1$.

For the PCS when $t=0$ the squeezed regions in the $(r,\Phi)$-plane
are symmetric around specific lines on the $\Phi$-axis and there is
a direct relation between $F_{x}$ and $F_{y}$. Also maximum
squeezing occurs at the most inner contours.  However, when the interaction is switched
on the features of the initial squeezed regions are changed and
 the following  results have been obtained:\newline
\noindent(i) For weak coupling  and when $\omega$ is much greater than $\lambda$
the initial squeezed regions move along the $\Phi$-axis by an amount
$2\omega t$.
Nevertheless, when $\omega$ is greater than $\lambda$ but not too much
the size of the squeezed regions are decreased compared to those of the initial ones.\newline
\noindent(ii)
For strong coupling  $\lambda >>\omega$ the initial inherited squeezed regions
for the system are lost, whereas when
$\lambda$ is  greater than  $\omega$ but not too much
the area of the squeezed regions are decreased compared to those of the initial
ones. Further,  the squeezed region associated with $F_{y}$ is decreased faster
than that with $F_{x}$.\newline
\noindent(iii) For the resonance case $\omega=\lambda$,
the variances associated with the $\hat{k}_{x}$- and $\hat{k}_{y}$-components
become polynomials of the normalized time. In this case  $F_{y}$ exhibits steady-state
squeezing, whereas the initial squeezing inherited in $F_{x}$ is completely suppressed
as the time evolves.

On the other hand, the BGCS is a minimum-uncertainty state, however, as result of the interaction
 with the nonlinear medium squeezing can be generated.
The results related to this case has been restricted to
two limiting cases, namely,
weak intensity case $|Z|<<1$ and strong intensity case $|Z|>>1$, and
 can be summarized as follows:

\noindent(i) For both weak  coupling and weak intensity maximum squeezing
in the $(|Z|,\Phi)$-plane occurs in the $F_{x}$ and $F_{y}$
at the most inner contours and on the line $\Phi=7\pi/4$, respectively.
However, in the strong intensity regime $F_{j}, j=x,y$ become
$|Z|$-independent and the squeezed area for $F_{x}(F_{y})$
decreases (increases) as $|Z|$ increases.

\noindent(ii) For strong coupling  squeezing occurs for short time interaction
only in the $F_{x}$, which exhibits single-peak structure in the $(|Z|,\Phi)$-plane.
Maximum value of squeezing occurs at $\Phi=0,2\pi$ and as $|Z|$ increases
$F_{x}$ gradually becomes $|Z|$-independent.

\noindent(iii) For the resonance  case
the variances associated with the $\hat{k}_{x}$- and $\hat{k}_{y}$-components
become polynomials of the normalized time, as in the PCS case.  For this case squeezing
have been detected for short interaction time in both quadratures
and  the behaviour of the $F_{x}(F_{y})$ is almost as that of  $F_{y}(F_{x})$
for the weak coupling  case.

\section*{References}

\end{document}